# Signatures of Chiral Anomaly in the Magnetoresistance of a Quasi-3-Dimensional Electron Gas at the Interface of LaVO$_3$ and KTaO$_3$

Harsha Silotia,[1] Anamika Kumari,[1] Anshu Gupta[1], Joydip De,[2] Santanu Kumar Pal,[2] Ruchi Tomar,[1] and S. Chakraverty[1, *]

1. *Quantum Materials and Devices Unit, Institute of Nano Science and Technology, Sector-81, Punjab, 140306, India.*
2. *Indian Institute of Science Education and Research Mohali Knowledge city, Sector 81, S.A.S. Nagar, Manauli, 140306, India.*

   *Email ID: suvankar.chakraverty@gmail.com*

**Abstract**: In a Dirac semimetal charges flow between two Weyl nodes when electric and magnetic fields ($B||E$) are parallel to each other manifesting interesting physical properties such as negative longitudinal magnetoresistance, planar Hall effect and anisotropic magnetoresistance. We observe a co-existence of weak antilocalization with large negative longitudinal magnetoresistance and an unusual Hall resistance with ($B||E$) configuration, at the conducting interface of LaVO$_3$ and KTaO$_3$. The depth of the conducting channel at the interface is estimated to be around 30 nm by using spectroscopy techniques of photoluminescence and time-correlated single-photon counting. Both planar Hall effect and anisotropic magnetoresistance exhibit oscillatory behaviour as a function of the angle between $E$ and $B$. A very similar temperature dependence of negative longitudinal magnetoresistance, planar Hall effect and anisotropic magnetoresistance suggest a strong correlation among them.

In recent years nontrivial topological properties of the matter have drawn enormous interest in the field of condensed matter physics. The main important reason is that such materials are not only interesting from the point of view of technological applications but also demonstrate rich fundamental physics. One such important example is the introduction of the Weyl equation in the context of Weyl semimetals (WSM) that was originally implemented in high-energy physics. In WSM, two Dirac nodes (the crossing point of two non-degenerate bands) are separated in momentum space generating two Weyl nodes [1-3]. Each Weyl node is characterized by a definite handedness known as chirality. The massless fermions with left or right-handed (spins rotated clockwise or anticlockwise around the node in momentum space) chirality located near two Weyl nodes are immiscible. Nonetheless, when electric and magnetic fields are applied parallel to each other they mix and the number of Weyl fermions with specific chirality no more remains conserved. This can be viewed as the creation of Weyl fermions at one nodal point with some chirality and annihilation of Weyl fermions at the other node with opposite chirality, this produces a current known as "axial current". This phenomenon is known as the Adler-Bell-Jackiw (ABJ) anomaly [4, 5]. Emergent physical properties such as negative longitudinal magnetoresistance (NLMR) and *topological* planar Hall effect (PHE) had been observed as a consequence of the ABJ anomaly [6].

Perovskite oxides are a class of materials that host an enormous amount of rich physical properties [7-18]. In addition, due to their simple cubic structures, heterostructures of different perovskite oxides can rather easily be prepared. Realization of Weyl fermions in perovskite oxides may give additional freedom of integrating other physical properties to achieve higher functionalities and emergent physical properties if appropriate heterostructure is designed. So far there is no report of the observation of ABJ anomaly and Weyl fermions in perovskite oxide materials. In this letter, we show the signature of the ABJ anomaly and the possible existence of (quasi) 3-dimensional Weyl fermions at the conducting interface of LaVO$_3$ (LVO) and KTaO$_3$ (KTO), by using magneto-transport and optical spectroscopic measurements. A large NLMR and *topological* PHE have been observed when an electric and magnetic field is applied parallel to each other. A larger enhancement in axial relaxation time ($\tau_a$) (relaxation time related to axial current) is observed. We have performed temperature dependent magnetoresistance and PHE measurements to find the correlation between them. The depth of the conducting channel is estimated from optical spectroscopy measurements.

Seven monolayers of crystalline LVO film were grown on KTO (001) single crystal substrate by using a pulsed laser deposition system. The substrate temperature was set at 600°C during the growth with the oxygen partial pressure of 1x10$^{-6}$ Torr. After deposition, cooling was done at the rate of 20°C/min in the same oxygen partial pressure. The electrical and magneto-transport properties were measured with the help of Quantum design physical property measurement system (PPMS). Contacts at the interface were made with the help of ultrasonic wire bonder in Hall geometry. The current is applied parallel to the plane (001) and the angle between the current and the magnetic field was varied using the horizontal rotator of PPMS. The depth of the conducting channel was measured at the interface using spectroscopy techniques of photoluminescence (PL) and time-correlated single-photon counting (TCSPC).

Figure 1(a) represents the resistivity of LVO/KTO interface as a function of temperature. A monotonic decrease in resistivity with decrease in temperature is observed which confirms the formation of a conducting channel at the interface of LVO/KTO. Magneto-transport measurements were performed by applying the magnetic field parallel to the current direction ($B||I$), i.e., the angle ($\theta$) between the magnetic field and current is kept zero. The measurement geometry is shown in the inset of Fig. 1(a). Figure 1(b) shows the change in the longitudinal magnetoconductivity (conductance along the direction of current) as a function of the applied magnetic field at selected temperatures. The longitudinal magnetoconductivity shows positive behaviour, with the increase in the magnetic field the conductivity increases (a NLMR). A significant NLMR observed upto 70K, above which it becomes vanishingly small. This large positive longitudinal magnetoconductivity or NLMR for $B||I$ configuration is regarded as an evidence of chiral anomaly (Weyl Fermions) [19-20]. In chiral anomaly, a charge pumping occurs between two Weyl nodes with opposite chirality in the presence of the parallel magnetic field and the electric field direction. This creates a charge imbalance between two nodes which can be equilibrated through an internode scattering process [21, 22]. Since this internode transfer process is related to a large momentum transfer and hence is rather weak. This causes an increase in NLMR, and can be explained using semi-classical Boltzmann equation. The conductivity tensor for chiral anomaly in the semi-classical limit is given by [20, 23],

$$\sigma_{xx} = \frac{e^2}{4\pi^2 \hbar c} \left(\frac{v}{c}\right) \frac{(evB)^2}{E_f^2} \tau_a \qquad (1)$$

Where, $\tau_a$ is the axial charge relaxation time arising due to the axial current as a consequence of chiral anomaly, e, ℏ, c, $E_f$, $v$ is the electronic charge, reduced Planck constant, speed of light, Fermi energy and velocity of electron, respectively.

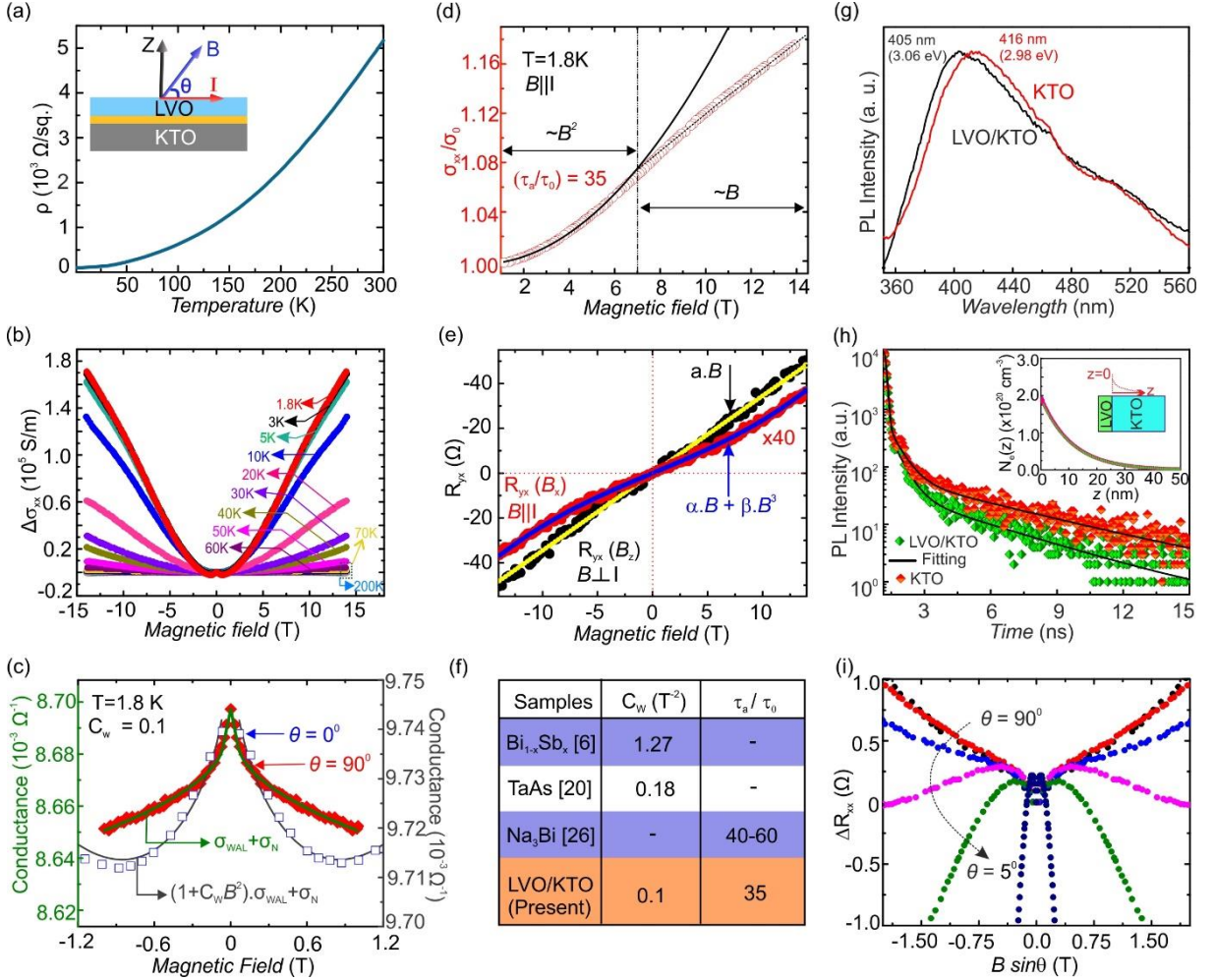

Fig.1. (a) Temperature dependent resistivity of LVO/KTO heterostructure (Schematic shown in inset). (b) The magnetoconductivity ($\Delta\sigma_{xx}$) as a function of magnetic field at various temperatures. (c) LMC [open blue square for $\theta = 0^o$ and red square for $\theta = 90^o$] at 1.8 K as a function of magnetic field showing WAL because of significant SOC. The dark grey and green curves show the fitting using equation $(1 + C_w B^2)\sigma_{WAL} + \sigma_N$ and $\sigma_{WAL} + \sigma_N$, respectively. (d) The dependence of conductivity ratio $\frac{\sigma_{xx}}{\sigma_0}$ with magnetic field for $B||I$ geometry. (e) The Hall resistance $R_{yx}$ as a function of magnetic field for $B||I$ [$R_{yx}(B_x)$] and $B \perp I$ [$R_{yx}(B_z)$]. The yellow and blue lines showing the fitting of $R_{yx}(B_z)$ and $R_{yx}(B_x)$ by using $a.B$ and $\alpha.B + \beta.B^3$, respectively. (f) Comparison of different parameters of various samples reported with LVO/KTO heterostructure. (g) Photoluminescence (PL) of bare KTO (001) substrate and LVO/KTO heterostructure. (h) The PL decay dynamics of KTO (001) substrate and LVO/KTO heterostructure at excitation wavelength of 375 nm and measured at 412 nm and 403 nm, respectively (the inset showing the carrier density distribution perpendicular to interface as the function of distance ($z = 0$ at interface) towards KTO (001) substrate as shown in the schematic also). (i) Change in resistance as a function of the perpendicular component of magnetic field $B \sin\theta$ for different $\theta$.

The LMR is measured along the direction of current by varying the angle between the current and the magnetic field direction. When the magnetic field is applied perpendicular to the current direction ($\theta = 90^0$ : $B \perp I$), a positive (negative) magnetoresistance (magnetoconductance) is observed, i.e., with increasing magnetic field resistance (conductance) increases (decreases). Figure 1(c) shows the experimental data of longitudinal magnetoconductance (LMC) for $B \perp I$ and $B||I$ configurations, presented as solid red and open blue squares, respectively. For $B \perp I$, LMC has a sharp peak at $B = 0$ T due to the weak-antilocalization and with increasing magnetic field, the conductance decreases monotonically in low magnetic field region. This LMC is well reproduced by the equation $\sigma_{WAL} + \sigma_N$, where $\sigma_{WAL}$ is the conductivity from weak antilocalization and $\sigma_N$ is from conventional Fermi surface contributions. The green line represents the fitted curve. Surprisingly, for $B||I$ configuration, in the low magnetic field the conductivity is showing a dip near 0.6 T and then it increases and there is peak at $B = 0$ T. The decrease in conductance at low magnetic field is signature of weak antilocalization. This coexistence of weak antilocalization and increasing conductivity (NLMR) is remarkable. This feature could be attributed to the chiral anomaly in presence of a strong spin orbit coupling. Similar to the previous reports our experimental LMC data for $B||I$ is also fitted excellently by the theoretical equation [6, 20].

$$\sigma_{xx}(B) = (1 + C_w B^2) \cdot \sigma_{WAL} + \sigma_N \qquad (2)$$

where,

$$\sigma_{WAL} = \sigma_0 + a_0 \sqrt{B} \qquad (3)$$

and,

$$\sigma_N^{-1} = \rho_0 + A_0 \cdot B^2 \qquad (4)$$

$\sigma_0$ is the zero field conductivity and $\rho_0$, $A_0$, $a_0$ are constants. The most important feature of this equation is the appearance of $C_w B^2$ (where $C_w$ is the constant and ~ 0.1 in our case) this term has a pure topological origin [6, 24]. Such topological term generates a chiral current known as axial current when $B||I$ configuration.

For further analysis, in Fig. 1(d) we have plotted the magnetoconductance as a function of applied magnetic field in +1 T to +14 T range. This figure suggests that upto ~ 7 T the magnetoconductance has a $B^2$ dependence and above that it has a $B$ linear dependence: a transition from semi-classical to quantum limit [25, 26]. We have then fitted the experimental data with Eq. (1) upto 7 T. The Eq. (1) provides an excellent fit for the experimental data $\frac{\sigma_{xx}}{\sigma_0}$, for temperature 1.8 K the extracted $\tau_a$ is ~ 35 times larger than the Drude relaxation time ($\tau_0$) of the system [26]. Figure 1(e) shows the transverse resistance data (resistance measured perpendicular to the direction of current) as a function of applied magnetic field for both out of plane ($B \perp I$) and $B||I$ geometry. The out of plane measurements shows the conventional Hall resistance R[yx] ($B_z$) having the linear dependence on $B$, fitted by $a.B$ equation (yellow curve). This suggests the absence of multi-channel conductance as well as anomalous Hall effect. The most surprising observation is the appearance of a non-linear (Hall) transverse voltage in $B||I$ configuration. This cannot be explained in the frame work of the conventional magneto-transport theory. Previous report suggest that such Hall effect has a pure topological origin, which originates from 3D Weyl fermions [6]. In this respect this Hall effect can be regarded as *topological* Hall effect. In contrast to the conventional Hall effect, $B||I$ Hall resistance R[yx] ($B_x$) has both linear and cubic contributions [6]. The R[yx] ($B_x$) data, shown by red curve have been magnified by 40 times to compare with the conventional Hall data. The blue line is the fitted curve of $R_{yx}$ ($B_x$) by the equation $\alpha.B + \beta.B^3$ where α and β were found to be 0.052 and 7.55 ×10$^{-5}$ respectively. We have presented the table of fitting parameters for different reported systems and compared them with for our LVO/KTO system in Fig. 1(i). Where, $C_w$ is extracted by fitting the LMC for $B||I$ by using Eq. (2) and $\frac{\tau_a}{\tau_0}$ is extracted by fitting the conductivity tensor $\frac{\sigma_{xx}}{\sigma_0}$ by using Eq. (1).

To estimate the thickness of the conducting channel at the interface we have done PL and TCSPC measurements. Figure 1(g) shows the PL of the single crystal KTO (001) and LVO/KTO interface. The PL peak of KTO (001) substrate is at 2.98 eV and for LVO/KTO heterostructure the peak is at 3.06 eV, the blue PL of the LVO/KTO interface suggests that the system is electron doped. The resulting PL is due to the radiative recombination of the photo generated holes and the doped electrons and hence can provide an estimate of the carrier density and their distribution at the interface of LVO/KTO [27]. Figure 1(h) presents the PL decay dynamics of the carriers at the interface and suggests that the PL decay dynamics is dominated by the non-radiative Auger recombination [27-28]. The temporal variation of PL for large spatial distribution of carriers is expressed as [28-30],

$$I(t) \propto B_m \int_0^\infty N_e(z)p(z)dz \qquad (5)$$

Where, $I(t)$ is the intensity of PL, $B_m$ is coefficient of bimolecular radiative recombination, $N_e(z)$ is the doped carrier density of the electrons at the interface, $p(z)$ is the photogenerated holes, and $z$ is the axis perpendicular to the interface (($z=0$) at the interface and $z>0$ is going towards the KTO substrate side as shown in the inset of Fig. 1(h)). From our previous report the solution of Eq. (5) is given by [28],

$$I(t) \propto \frac{K}{\delta}e^{-At}\left[1 - erf\left(N_o\sqrt{C}t\right)\right] \qquad (6)$$

Where, $N_o\sqrt{C}t$ is the argument of error function, $K$ is the constant, $\frac{1}{\delta}$ is defining the depth of the quantum well formed at the interface, $A$ is the coefficient of single photocarrier trapping, $C$ is the coefficient of Auger recombination, $N_o$ is the density of electron at $z=0$ and t is the time. By using Eq. (6) we have fitted the PL decay curve and from that we get the carrier density at the interface is ~ 1.95x10$^{20}$ cm$^{-3}$, which is very closed to the carrier density calculated from the Hall effect that is ~ 1.75x10$^{20}$ cm$^{-3}$ and the depth of quantum well is ~ 30 nm. Hence this conducting channel could be considered as quasi-3-dimensional. For further confirmation of the channel dimensionality, we have checked if the scaling properties of magnetoresistance is satisfied for different angles or not. To check this scaling, we have plotted the change in the longitudinal resistance as a function of perpendicular component of magnetic field $Bsin\theta$ for different $\theta$ values. For a two dimensional conducting channel only the perpendicular component of the magnetic field to the plane should matter. But the Fig. 1(i) clearly suggest that this is not the case for our system that also indicates a 3-dimensional nature of the conducting channel.

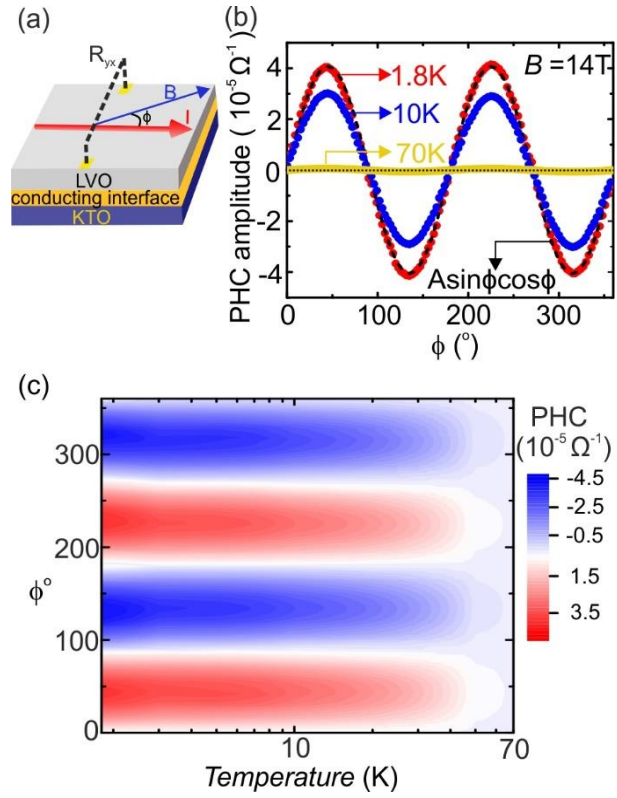

Fig.2. (a) Schematic geometry for Planar Hall effect (PHE) measurements. (b) Angle dependent PHC measured at different temperatures for 14 T (The black dash line is the fitted curve). (c) In plane angle ($\phi$) and temperature dependent contour plot for PHC.

Figure 2(a) is the schematic illustration of the electrical measurements of the transverse resistance ($R_{yx}$) defined as the PHE. In these measurements, both the magnetic field and the current are applied in the same plane defined by the interface of LVO and KTO. The resistance $R_{yx}$ is measured perpendicular to the direction of the current. The in-plane angle between the current and the applied magnetic field is defined as $\phi$. The planar Hall conductance (PHC) is recorded as a function of $\phi$ by rotating the magnetic field direction with respect to the current, keeping the magnetic field strength constant. The PHC shows a periodic behavior with two-fold oscillations as a function of $\phi$, with maxima at $45^0$ and minima at $135^0$ as shown in Fig. 2(b). [31-36]. The black dotted curve (fitted) shows $\sin\phi \cos\phi$ dependence of the two-fold oscillations. This angular dependence is very similar to what observed in topological materials [32, 37]. This angular dependence of PHC for 14 T field is measured at different temperatures. Figure 2(c) represents the contour plot of the PHC in temperature - $\phi$ space. With increase in temperature, the amplitude of PHC oscillations decreases and vanishes at ~ 80 K. Figure 3(a) shows the amplitude of PHC as a function of temperature. Amplitude is extracted by fitting PHC data by the equation, $\sigma_{yx}^{PHC} = A \sin\phi \cos\phi$. In addition, we have also plotted the temperature dependent ratio of the axial charge relaxation time ($\tau_a$) to Drude relaxation time ($\tau_0$) as shown in Fig. 3(b). This ratio decreases with increasing temperature. The long relaxation time ($\tau_a$) for the axial current is due to presence of chiral anomaly.

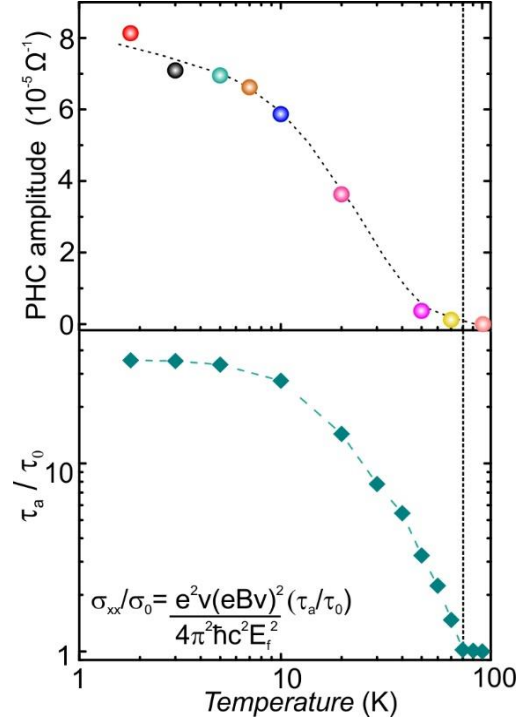

Fig.3. (a) Variation of PHC amplitude (solid circles) with temperature. (b) Variation of the ratio of axial charge relaxation time ($\tau_a$) and Drude relaxation time ($\tau_0$) with temperature.

At 80 K, the ratio $\frac{\tau_a}{\tau_0}$ becomes unity which signifies the disappearance of chiral anomaly at high temperatures. The amplitude of PHC and the ratio $\frac{\tau_a}{\tau_0}$ both vanishes at the same temperature, indicating that both PHC and NLMR has same physical origin.

In summary, we observed a large NLMR and *topological* PHE at the conducting interface of LVO and KTO, when $E$ is parallel to $B$. In the low magnetic field NLMR follows a $B^2$ dependence and above 7 T it becomes $B$ linear. This is in agreement with chiral anomaly theory: a transition from semi-classical to quantum limit. The Hall resistance has a $\alpha.B + \beta.B^3$ dependence in $E||B$ configuration. Such magnetic field dependence suggests that this Hall effect originated from 3-dimensional Weyl fermions. These two observations are consistent with each other and both point towards the chiral anomaly theory arising from 3-dimensional Weyl fermions. The optical spectroscopy data further strengthen this, it suggests a finite width of the conducting channel: a quasi-3D conducting system. In addition, our result provides an estimate for $\tau_a$ which is enhanced by a factor of ~ 35 at 1.8 K in compare to the Drude conductivity of the system, in $E||B$ configuration. Both axial current and PHE vanish at ~ 80 K suggesting they have the same physical origin. Our result may provide an unexplored pathway to design new topological materials based on oxide interfaces. Our observations demand detailed theory and further spectroscopic measurements of this system. In addition, a relatively high-temperature observation of the PHE and negative magnetoresistance suggest an exciting possibility of creating a new vector magnetic field sensor and magnetic memory devices based on similar materials.

**ACKNOWLEDGEMENTS:**

This project was supported by Grant No. 58/14/17/2019-BRNS/37024(BRNS).

Conflict of Interest: The authors declare no conflict of interest.

References:

[1] X. Wan, A.M. Turner, A. Vishwanath, and S. Y. Savrasov, Phys. Rev. B 83, 205101 (2011).

[2] G. B. Halasz, and L. Balents, Phys. Rev. B 85, 035103 (2012).

[3] G. Xu, H. Weng, Z. Wang, X. Dai, and Z. Fang, Phys. Rev. Lett. 107, 186806 (2011).

[4] S. L. Adler, Phys. Rev. 177, 2426 (1969).

[5] J. S. Bell, and R. Jackiw, Nuovo Cimento A 60, 47 (1969).

[6] H. J. Kim, K. S. Kim, J. F. Wang, M. Sasaki, N. Satoh, A. Ohnishi, M. Kitaura, M. Yang, and L. Li, Phys. Rev. Lett. 111, 246603 (2013).

[7] N. Wadehra, R. Tomar, R. M. Varma, R. K. Gopal, Y. Singh, S. Dattagupta, and S. Chakraverty, Nat. Commun. 11, 1–7 (2020).

[8] S. Goyal, N. Wadehra, and S. Chakraverty, Adv. Mater. Interfaces 7, 2000646 (2020).

[9] M. Dumen, A. Singh, S. Goyal, C. Bera, and S. Chakraverty, J. Phys. Chem. C 125, 15510–15515 (2021).

[10] H. Y. Hwang, Y. Iwasa, M. Kawasaki, B. Keimer, N. Nagaosa, and Y. Tokura, Nat. Mater. 11, 103–113 (2012).

[11] A. Ohtomo, H. Y. Hwang, Nature 427, 423–426 (2004).

[12] H. Zhang, Y. Yun, X. Zhang, H. Zhang, Y. Ma, X. Yan, F. Wang, G. Li, R. Li, T. Khan, C. Yuansha, W. Liu, F. Hu, B. Liu, B. Shen, W. Han, and J. Sun, Phys. Rev. Lett. 121, 116803 (2018).

[13] N. Kumar, N. Wadehra, R. Tomar, S. Kumar, Y. Singh, S. Dattagupta, and S. Chakraverty, Adv. Quant. Tech. 4, 2000081 (2021).

[14] C. Liu, X. Yan, D. Jin, Y. Ma, H. W. Hsiao, Y. Lin, T. M. Bretz-Sullivan, X. Zhou, J. Pearson, B. Fisher, J. S. Jiang, W. Han, J. M. Zuo, J. Wen, D. D. Fong, J. Sun,

H. Zhou, and A. Bhattacharya, arXiv preprint arXiv:2004.07416 (2020).

[15] Z. Chen, Z. Liu, Y. Sun, X. Chen, Y. Liu, H. Zhang, H. Li, M. Zhang, S. Hong, T. Ren, C. Zhang, H. Tian, Y. Zhou, J. Sun, and Y. Xie, Phys. Rev. Lett. 126, 026802 (2021).

[16] Z. Chen, Y. LiU, H. Zhang, Z. Liu, H. Tian, Y. Sun, M. Zhang, Y. Zhou, J. Sun, and Y. Xie, Science 372, 721–724 (2021).

[17] N. Wadehra, and S. Chakraverty, App. Phys. Lett. 114, 163103 (2019).

[18] R. Tomar, S. Kakkar, C. Bera, and S. Chakraverty, Phys. Rev. B 103, 115407 (2021).

[19] N.P. Ong, and S. Liang, Nat. Rev. Phys. 3, 394–404 (2021).

[20] X. Huang, L. Zhao, Y. Long, P. Wang, D. Chen, Z. Yang, H. Liang, M. Xue, H. Weng, Z. Fang, X. Dai, and G. Chen, Phys. Rev. X 5, 031023 (2015).

[21] H. B. Nielsen, and M. Ninomiya, Phys. Lett. B 105, 219223 (1981).

[22] H. B. Nielsen, and M. Ninomiya, Phys. Lett. B 130, 389396 (1983).

[23] D. T. Son, and B. Z. Spivak, Phys. Rev. B 88, 104412 (2013).

[24] N. W. Ashcroft, and N. D. Mermin, Solid State Physics (Saunders, New York, 1976), Chap. 13.

[25] K. Das, S. K. Singh, and A. Agarwal, Phys. Rev. Research 2, 033511 (2020).

[26] J. Xiong, S. K. Kushwaha, T. Liang, J. W. Krizan, M. Hirschberger, W. Wang, R. J. Cava, and N. P. Ong, Science 350, 413–416 (2015).

[27] A. Kumari, J. De, S. Dattagupta, H. N. Ghosh, S. K. Pal, and S. Chakraverty, Phys. Rev. B 104, L081111 (2021).

[28] Y. Yamada, H. K. Sato, Y. Hikita, H. Y. Hwang, and Y. Kanemitsu, Appl. Phys. Lett. 104, 151907 (2014).

[29] H. Yasuda, Y. Yamada, T. Tayagaki, and Y. Kanemitsu, Phys. Rev. B 78, 233202 (2008).

[30] Y. Yamada, H. Yasuda, T. Tayagaki, and Y. Kanemitsu, Appl. Phys. Lett. 95, 121112 (2009).

[31] A. A. Taskin, H. F. Legg, F. Yang, S. Sasaki, Y. Kanai, K. Matsumoto, A. Rosch, and Y. Ando, Nat. Commun. 8, 1–7 (2017).

[32] J. Meng, H. Xue, M. Liu, W. Jiang, Z. Zhang, J. Ling, L. He, R. Dou, C. Xiong, and

J. Nie, J. Phys.Condens. Matter 32, 015702 (2019).

[33] S. Nandy, G. Sharma, A. Taraphder, and S. Tewari, Phys. Rev. Lett. 119, 176804 (2017).

[34] N. Kumar, S. N. Guin, C. Felser, and C. Shekhar, Phys. Rev. B 98, 041103 (2018).

[35] D. D. Liang, Y. J. Wang, W. L. Zhen, J. Yang, S. R. Weng, X. Yan, Y. Y. Han, W. Tong, W. K. Zhu, L. Pi, and C. J. Zhang, AIP Adv. 9, 055015 (2019).

[36] R. Singha, S. Roy, A. Pariari, B. Satpati, and P. Mandal, Phys. Rev. B 98, 081103 (2018).

[37] H. Li, H. W. Wang, H. He, J. Wang, and S. Q. Shen, Phys. Rev. B 97, 201110 (2018).